\documentstyle[12pt,aasms4]{article}
\accepted{18 March 1999}

\newcommand\nion[2]{#1\,\lowercase{{\sc #2}}}
\newcommand\wave[1]{\mbox{$\lambda$#1\,\AA}}

\def\BmV0{\mbox{(B-V)$^{\rm o}$}}
\def\VmK0{\mbox{(V-K)$^{\rm o}$}}
\def\MV0{\mbox{M$_{\rm V}^{\rm o}$}}

\def\carbiso{\mbox{${\rm ^{12}C/^{13}C}$}}

\lefthead{}
\righthead{}
\begin{document}

\title{ 
R-Process Abundances and Chronometers in Metal-Poor Stars}

\author{John J. Cowan,\altaffilmark{1}
B. Pfeiffer,\altaffilmark{2}
K.-L. Kratz,\altaffilmark{2}
F.-K. Thielemann,\altaffilmark{3}
Christopher Sneden,\altaffilmark{4}
Scott Burles,\altaffilmark{5,6}
David Tytler,\altaffilmark{5} and Timothy C. Beers\altaffilmark{7}}
\begin {center}
cowan@mail.nhn.ou.edu,
pfeiffer@vkcmzd.chemie.uni-mainz.de,
klkratz@vkcmzd.chemie.uni-mainz.de, 
fkt@quasar.physik.unibas.ch,
chris@verdi.as.utexas.edu, scott@rigoletto.uchicago.edu, tytler@ucsd.edu,
beers@pa.msu.edu

\end {center}

\vskip .5truein
\begin{center}
To appear in {\it The Astrophysical Journal}, {\bf 521}
\end{center}

\altaffiltext{1}{Department of Physics and Astronomy,
University of Oklahoma, Norman, OK 73019 }
\altaffiltext{2}{Institut f\"ur Kernchemie, Univ. Mainz, Fritz-Strassmann-
Weg 2, D-55099 Mainz, Germany}
\altaffiltext{3}{Department f\"ur Physik und Astronomie, Univ. Basel,
Klingelbergstr. 82, CH-4056 Basel, Switzerland}
\altaffiltext{4}{
Department of Astronomy and McDonald
 Observatory, University of Texas, Austin, TX 78712}
\altaffiltext{5}{Department of Physics and Center for Astrophysics and Space
Sciences, University of California, San Diego, MS 0424, La Jolla, CA
92093-0424}
\altaffiltext{6}{Department of Astronomy and Astrophysics,
University of Chicago, Chicago, IL 60637}
\altaffiltext{7}{Department of Physics and Astronomy,
Michigan State University, E. Lansing, MI 48824}

\begin{abstract}

Rapid neutron-capture (i.e., {\it r}-process) 
nucleosynthesis calculations, employing internally consistent
and physically realistic nuclear physics input  (QRPA $\beta$-decay properties
and
the recent ETFSI-Q nuclear mass model),
have been  performed.
These theoretical computations  assume 
the classical waiting-point approximation 
of 
(n,$\gamma$) ${\rightarrow \atop \leftarrow}$ ($\gamma$,n) equilibrium.
The calculations  
reproduce the solar isotopic $r$-abundances in detail, including the 
heaviest stable Pb and Bi isotopes. 
These calculations are then compared with 
ground-based and HST observations of  neutron-capture  
elements in the metal-poor halo stars CS~22892--052, HD~115444, HD~122563 
and HD~126238. 
The elemental abundances in all four  metal-poor stars are consistent
with the solar {\it r}-process elemental distribution for the elements 
Z $\ge$ 56. These results strongly suggest, at least for those elements, 
that the relative elemental {\it r}-process abundances have not changed 
over the history of the Galaxy. 
This indicates also that 
it is unlikely that the solar {\it r}-process abundances resulted  
from a random superposition of varying abundance patterns from 
different {\it r}-process nucleosynthesis sites.
This further  suggests that there is  one 
{\it r}-process site in the Galaxy, at least for 
elements Z $\ge$ 56. 

Employing the observed stellar abundances of stable elements, 
in conjunction with the 
solar {\it r}-process abundances
to constrain the calculations, 
predictions for the zero decay-age abundances  of the radioactive elements 
Th and U are presented.
We compare these predictions (obtained with the mass model ETFSI-Q, which 
reproduces solar $r$-abundances best) with newly derived 
observational values in three very  metal-poor halo stars:
HD 115444, CS~22892--052 and HD 122563. 
Within the observational errors the ratio of 
[Th/Eu] is the same in both CS~22892--052 and HD~115444.
Comparing with the theoretical ratio suggests an 
average age of these two very metal-poor stars to be 
15.6 $\pm$ 4.6 Gyr, consistent with 
earlier radioactive age estimates and recent globular and 
cosmological age estimates.
Our upper limit on the uranium abundance in HD 115444 also
implies a similar age.
Such radioactive age determinations of very low metallicity stars 
avoid uncertainties in Galactic chemical evolution models.
They still include uncertainties due to the involved nuclear 
physics far from beta-stability. However, we give an extensive
overview of the possible variations expected and come to the conclusion
that this aspect alone should not exceed limits of 3 Gyr.
Therefore this method shows
promise as an independent dating technique for the Galaxy. 

\end{abstract}


\section{Introduction}

Elemental  abundances in metal-poor halo stars provide important clues
to the chemical evolution of the early Galaxy.
Since these stars are extremely old, their abundance patterns are
living records of the first generations of Galactic nucleosynthesis.
Much recent observational and theoretical work has concentrated on
understanding the origin of neutron-capture elements in halo stars.
These elements are those with atomic numbers Z~$>$~30 (or mass numbers 
A~$>$~68), thus comprising the majority of elements in the 
periodic table.
Some isotopes of the neutron-capture elements may be built
through neutron bombardment rates that are slow enough to allow all 
beta-decays to take place between successive neutron captures; this
is called the  {\it s}-process.
Alternatively, short-duration rapid neutron bombardments that occur
more rapidly than 
beta-decay rates can produce other neutron-capture element isotopes;
this is the {\it r}-process.
More generally, many isotopes may arise from  a 
combination of both {\it s}- and {\it r}-processes,
and great care must be excercized in determining 
the relative contributions of each of these
processes to the solar and stellar elemental abundances.

Many chemical composition studies of low-metallicity stars have
detected the presence of a strong {\it r}-process component in the neutron
capture
element abundances  
(\cite{spi78,tru81,sne83,sne85,gil88},
Gratton \& Sneden 1991, 1994; \cite {sne94}; McWilliam 
et al. 1995a, 1995b; \cite{cow96,sne96,bur98}; McWilliam 1998).
In addition, there has been increasing evidence suggesting a relative or
scaled solar 
{\it r}-process abundance pattern in these stars, at least for the heaviest of
these elements  
(\cite{gil88,cow95,sne96,cow97}). 

Even though the 
elemental abundance 
 patterns apparently 
are in agreement with a purely
solar {\it r}-process distribution,  these are necessarily 
sums over individual isotopic
abundances. 
Therefore a successful  solar {\it r}-process match to stellar elemental 
abundances 
still gives fewer  constraints to their isotopic
composition than would a direct observation of a solar isotopic 
{\it r}-process
abundance pattern. In principle, there exists a degeneracy in the
sense that different non-solar isotopic abundance patterns might still fit
a solar {\it r}-process elemental pattern. 
It is theoretically possible to produce identical abundances for 
approximately 
30 elements with Z~$\ge$~56 through different mixtures of the approximately
80 individual isotopes of these elements.
It is harder to do this with superpositions from different r-process
calculations.
Goriely \& Arnould (1997) have shown that 
a series of several non-solar isotopic abundance distributions
could produce a total elemental abundance pattern that indeed matches
closely some of the neutron-capture elements in one low-metallicity star,
especially those in the more insensitive relatively flat regions in between
$r$-process peaks. However, if also the peaks have to be reproduced at their
precise locations (and there exists an increasing amount of data in the Pt
peak), this agreement can no longer  occur as a result of 
any random superposition of abundances.
There has to be an intriguing conspiracy for compensation of the non-solar
values of different isotopes in order to still fit a solar element
composition. The most reasonable explanation is that
the same set of superpositions occurs in each individual event. This would
also explain why such a pattern is already observed in the lowest metallicity
stars, where only one or at most a few events contributed.
Thus, while the observation of a solar {\it r}-process elemental
  pattern in certain mass regions is not an absolute proof for isotopic 
solar {\it r}-process abundances, it is
 the most reasonable and probable conclusion when also the peaks have to
be reproduced correctly.

This conclusion is further strengthened by the 
existence of a scaled  solar {\it r}-process abundance pattern
for many elements, 
not just in one 
low-metallicity star, but now in 
four stars (Sneden et al. 1996, 1998). 
The data for these stars are not restricted to a narrow region of
elements (i.e.,  the rather flat
inter-peak region near mass number 150), 
but now cover a wide elemental range from Ge (Z = 32) through 
Pb (Z = 82).
These recently observed abundance patterns also
agree with  earlier findings, for a wide range of stars 
with varying metallicities,  that the Ba/Eu ratio
varies from a pure {\it r}-process ratio at low metallicities to a solar mix 
at solar
metallicities 
(see McWilliam's 1997 review and references therein;
also McWilliam 1998, Burris et al. 1998).
Magain (1995) recently attempted to estimate Ba isotopic abundances from 
synthetic/observed profile comparisons of a Ba II line in the very 
low-metallicity subgiant HD~140283. 
He concluded that a total solar system isotopic mix (i.e.,
 {\it r}- + {\it s}-process) 
provided the best match, implying a significant {\it s}-process 
contribution to the
neutron-capture elements in this star.
However, Fran\c{c}ois \& Gacquer (1999) have repeated Magain's
investigation
with superior spectroscopic data, and derive an isotopic mix for this 
same star that is consistent with a purely {\it r}-process synthesis origin. 

There have been several difficulties, however, in determining the true nature
of the stellar abundance pattern.
First, most of the stellar abundance determinations already exist for the 
easily observed rare-earth elements, but these elements occur between
the 2$^{nd}$ and 3$^{rd}$ neutron-capture peaks, i.e., in 
the relatively flat inter-peak region.
The second {\it
r}-process peak
(near mass number, A, $\sim$ 130)  contains such elements as Te and Xe
that at present have no identifiable solar and stellar transitions. 
The 3$^{rd}$ solar peak (A~$\sim$~195, with elements Os, Ir and Pt) has
mostly 
UV transitions and is thus not accessible to ground-based observations. 

Additionally, 
there have been theoretical difficulties in predicting the neutron-capture
element
abundances. 
The first of these is the well-known uncertainty in the site of the {\it
r}-process 
(see Cowan, Thielemann, \& Truran 1991, \cite{mat92,mey94} for further
discussion of possible
sites). 
Even promising scenarios, like the high entropy neutrino wind in supernovae
(Takahashi, Witti, \& Janka  1994, Woosley et al. 1994, Qian \& Woosley 1996)
have problems in obtaining the required entropies,
and they have deficiencies in their abundance predictions (Freiburghaus et al.
1999, Meyer, McLaughlin, \& Fuller 1998). 
Thus, even though it would be  preferred, 
it is currently not possible to provide results
from a specific {\it r}-process site which give a good global fit to 
the observed, stable solar {\it r}-process abundances. 
Heretofore only observed stellar and scaled solar abundances have been 
compared. Therefore, 
at the very least 
we need to develop tools to theoretically predict ``zero-age'' solar 
{\it r}-process abundances for those long-lived unstable neutron-capture 
elements, where the solar abundances have been modified
by decay from the initial zero-age abundances produced in an r-process site.
The abundances of the stable elements in, and beyond, 
the 3$^{rd}$ {\it r}-process peak (up to A=209)
can be used to help constrain the predictions of such an initial
production pattern of the nuclei with
A~$>$~209, because the mass numbers 206, 207, 208, and 209 contain large
contributions from several alpha-decay chains of heavier nuclei. This helps
especially to predict the long-lived radioactive chronometers Th and U,
independent of knowing the site for the {\it r}-process.

A final problem  is that   
the nuclei involved in the {\it r}-process are far from stability 
and therefore the requisite nuclear data needed for 
reliable {\it r}-process predictions 
are in most cases not 
obtainable by experimental determination.
However, there have been recent advances in theoretical prescriptions for very
neutron-rich nuclear
data (M\"oller et al. 1995, Aboussir et al. 1995, 
M\"oller, Nix, \& Kratz 1997, Chen et al. 1995, Pearson, Nayak, Goriely 1996).

The  detection of thorium in the ultra-metal-poor halo star CS~22892-052
(\cite
{sne94,sne96,cow97})
in conjunction with the HST detections of the third {\it r}-process peak
elements in 
some halo stars
bears directly on nucleochronology.
The long-lived radioactive nuclei (known as chronometers)
in the thorium-uranium region
are formed entirely in the {\it r}-process and can be used to determine the
ages of
stars and the Galaxy. 
In other low-metallicity stars the detection of neutron-capture
elements in the (osmium-platinum)
3$^{rd}$ {\it r}-process peak,
a nuclear region near that of uranium and thorium,
confirms the full operation
of
the {\it r}-process for the synthesis of the heaviest
(Z~$>$~55) {\it r}-process elements
during the early history of the Galaxy.

The combination of the improved nuclear and stellar  data 
will now allow more comprehensive and detailed
attempts at a determination of the nature of the abundance patterns in these
earliest 
Galactic  stars. 
In this paper we analyze  the production of the {r}-process elements 
in the
solar system and in four metal-poor
halo stars.  
In \S 2 we describe  the {\it r}-process calculations
and their relevance to the solar system isotopic abundance mix.
Abundance comparisons between the calculations and the recent observations of 
four different metal-poor halo stars are made in  \S 3. 
Age estimates, based upon predicted and observed radioactive
chronometers in the stars CS 22892--052 and HD~115444 (and therefore also
lower limits for the Galactic age),
are given in \S 4. In \S 5 we conclude with 
a discussion of our results.

\section{Calculations}

We assume, as a working hypothesis, 
that the heavy element
abundances of very low metallicity stars are given by a pure {\it r}-process
composition. 
This assumption is supported 
by the 
observational evidence, at least for the elements beyond Ba, where data
are available. 
We have analyzed {\it r}-process abundances 
with predictions
from calculations in the waiting-point assumption (or 
(n,$\gamma$) ${\rightarrow \atop \leftarrow}$ ($\gamma$,n) equilibrium) 
with a continuously improving nuclear data base (Kratz et
al. 1988, Kratz et al. 1993, Thielemann et al. 1994, Chen et al. 1995,
Pfeiffer et al. 1997). This approach is based on a continuous superposition of
{\it r}-process components with a varying {\it r}-process 
path related to contour
lines of neutron separation energies in the range of 4 to 2 MeV; the
latter being determined by the combination of neutron number density and
temperature. 
Such a procedure is also site-independent and essentially meant to provide
a good fit to solar {\it r}-abundances, but it can also provide information
regarding the type of conditions that 
a ``real'' {\it r}-process site has to fulfill.
The fit is performed by adjusting the weight of the individual components
for different neutron separation energies S$_n$(n$_n$,T) and the time duration
$\tau$ for which these (constant) conditions are experienced, starting with
an initial abundance in the Fe-group. For a given (arbitrary) temperature,
$T$, $S_n$
is a function of neutron number density, $n_n$. 
The superposition weights $\omega$(n$_n$) and process
durations $\tau$(n$_n)$ have a behavior similar to powers of n$_n$. 
This corresponds to a linear relation in $\log(n_n)$ and is already observed 
when taking a minimum of three components in order to fit the A=80, 130, and 
195 abundance peaks (Kratz et al. 1993). 
This approach (although only a fit and not a realistic
site calculation) also has the advantage that such a continuous dependence
on physical conditions has embedded some relation to an (although unknown)
astrophysical site. Therefore, one expects some predictive power also for
mass regions that are not explicitly fitted. 

This technique should be preferable to a multi-event superposition 
of components with a permitted random relation between $\omega$, $\tau$ and
$n_n$ (Goriely \& Arnould 1996). In such a case, it is (similar to a Fourier
analysis) almost always possible to obtain a good fit for any type of nuclear 
mass model, but this is achieved by possibly even compensating for
deficiencies 
in nuclear properties. Such an approach is not random but its superpositions
can change randomly with different mass models.
By employing thousands of components, features within the fit interval can
be reproduced very well, but very probably - like for a high power polynomial 
fit - one expects less predictive power at the boundaries and outside of that 
fit region.  Such a procedure (solution to an inverse problem) is 
meaningful when essentially all physics is understood and only the
superposition
scheme has to be unraveled. This is the case for the s-process, involving
predominantly stable nuclei. It is thus comforting that Goriely (1997) could 
reproduce the long utilized exponential superposition of neutron exposures 
(see e.g. Seeger, Fowler, \& Clayton 1965; K\"appeler et al. 1989) for the 
$s$-process. This is a strong 
support for doing so also in $r$-process analyses, as long as we have to deal
with unknown uncertainties in nuclear physics.

Following our simplifying but physically predictive procedure, the best
agreement with solar system {\it r}-process abundances was obtained when
we employed the nuclear mass predictions from an extended Thomas Fermi
model with quenched shell effects far from stability (i.e., ETFSI-Q, 
Pearson, Nayak, \& Goriely 1996) and the beta-decay properties from 
QRPA-calculations based on the methods described in M\"oller \& Randrup 
(1990; see also M\"oller et al. 1997). The choice of this quenched 
mass formula is strengthened by recent experimental results for very 
neutron-rich isotopes in the $^{132}$Sn region (Kratz  1998, Fogelberg et
al. 1998). Also comparison of masses around N=82 are in better agreement with 
the ETFSI-Q predictions than with any other commonly used mass formula
(Isakov et al. 1998, Mach et al. 1998).

The necessary superposition of different components (with r-process paths
of a corresponding neutron separation energy $S_n$ and duration $\tau$)
can be expressed in terms of superposition weights $\omega(S_n)$ and durations
$\tau(S_n)$. How in the
still uncertain astrophysical environments such $S_n's$ are attained is
uncertain and not unique. For an $(n,\gamma)$-$(\gamma,n)$ equilibrium,
the maximum in each isotopic chain at a given $S_n$ can be attained for
arbitrary combinations of $n_n$ and $T$. For simplicity (but not uniqueness)
we have chosen $T=1.35\times 10^9$K, which makes $S_n(n_n,T)$ only a function
of $n_n$. Then the superposition can also be expressed in the form
$\omega(n_n)$ 
expressed in the form 
$\omega$(n$_n$) = 8.36 x 10$^{6}$n$_n^{-0.247}$ and
$\tau$(n$_n$) = 6.97 x 10$^{-2}$n$_n^{0.062}$ sec, when restricting the
fit to the mass regions around the r-process peaks (A=80, 130, 195),
where the paths come closest to stability, mass model extrapolations need
not be extended far into unknown territory, and even a limited amount of
experimental information is available. These power laws in $n_n$ play a 
similar role as an exponential superposition of neutron 
exposures in the classical s-process.
The (n,$\gamma$) ${\rightarrow \atop \leftarrow}$ ($\gamma$,n) equilibria
on which the {\it r}-process calculations are based,  will likely be 
obtained in any astrophysical environment with conditions in excess of 
n$_n$=$10^{20}$ cm$^{-3}$ and $T$~=~$10^9$K, as expected from all possible 
sites (Meyer 1994). The major remaining question is related to the 
assumption of an (n,$\gamma$) ${\rightarrow \atop \leftarrow}$ ($\gamma$,n)
equilibrium during the "freeze-out" phase in realistic astrophysical sites 
and depends on the temporal decline pattern of
neutron density and temperature below the
above mentioned limits. Thielemann et al. (1998) and Freiburghaus et al.
(1999) could show with full network calculations, and 
without an a priori assumption of
equilibria, that these freeze-out effects are small, when mass models which
include shell quenching far from stability are used.
They also showed that the smooth superposition of results from
different neutron separation energies with declining
weights for smaller S$_n$-values can be translated into an almost
constant entropy superposition of full network calculations
for masses with A~$>$~110 (see also Figure 2 and the discussion in 
Takahashi et al. 1994).  
The coefficients and powers were obtained from a least square fit to solar
{\it r-}abundances with the nuclear input discussed above.

We have used this physically motivated fitting procedure, 
which apparently reproduces
the stable solar {\it r}-process abundances very well with a simple and robust
superposition scheme, for the calculations in this paper.
The similarity to the well known and tested classical s-process
approach of an exponential superposition of neutron exposures provides a
further support for its application. It has the additional advantage
that it leads to clear predictions which have to and can be tested.
Therefore, in the present investigation we have extrapolated the calculations 
into unknown territory, specifically the actinides. 
Early applications of this scheme were only applied to fit abundances
up to the A~=~200, where individual isotopic data are
available. The challenge is to find some features which provide
a measure of the accuracy for all nuclei beyond the A~=~195 peak, 
including the
unstable actinides in alpha-decay chains (Thielemann et al. 1993).
In fact, there exist some observational constraints for these
mass regions as well. Up to A = 205, isotopic abundances stemming from 
the beta-decay
of {\it r}-process progenitor nuclei (as in the lighter mass regions) 
are available.
The $r$-abundances of the $^{206,207,208}$Pb isotopes and $^{209}$Bi
contain additional information and are dominated by contributions
from alpha-decay chains of isotopes with A $>$ 209. For $^{206}$Pb
and $^{207}$Pb, for example, 
$\beta$-decay from the {\it r}-progenitors 
    back to these isotopes amounts to only 5\%, whereas 95\% of the 
$r$-abundances come from alpha back-decays of trans-lead isotopes. 
    Similarly, for $^{208}$Pb and $^{209}$Bi the pure $\beta$ contribution is 
    about 15\% and 7\%.

In Figure~1 we show the results of our calculations before
and after considering beta-decay and
alpha-decay chains with short half-lives. 
The dashed and solid lines shown for A$>$206 make both use of the mass
model ETFSI-Q, but with slightly different superposition weights
$\omega(n_n)$ (discussed below) in order to indicate an uncertainty range
in the abundance predictions. For the nuclei $^{232}$Th, $^{235}$U, $^{238}$U,
and $^{244}$Pu this range is indicated by a vertical bar.
As demonstrated earlier by Pfeiffer et al. (1997),
the theoretical prescription described above provides predictions in 
excellent agreement with the observed solar abundances (small
filled circles in Figure 1), especially also for A~$>$~195.
This is also the case for $^{203}$Tl, $^{204}$Hg, $^{205}$Tl, as well as for 
the alpha-decay dominated Pb and Bi isotopes from A=206 to 209.
The $r$-contributions based on different $s$-process studies are listed
in the first part of Table 1. K\"appeler et al. (1989) made use of
the classical model of an exponential superposition of neutron exposures,
aiming for a best fit of pure $s$-nuclei. Beer et al. (1997) is based on 
stellar sites (thermal pulses with $^{13}$C and $^{22}$Ne neutron sources).
 
While we see appreciable variation between these two determinations
of $s$-process contributions (and therefore $r$-process residuals) for 
A=203, 204, and 205, we see a much larger variation in the results of
$r$-process calculations when utilizing different mass models.
Although we will later employ dominantly the ETFSI-Q mass model (for the
good reasons seen in Figure 1 and to be discussed below), we show in Table 1
predictions for $^{203}$Tl, $^{204}$Hg, $^{205}$Tl, $^{206-208}$Pb, and
$^{209}$Bi based on a large variety of mass formulas and models.
The mass region 203-205 is one of those locations before and after $r$-process
peaks, where too strong shell closures far from stability lead to
"abundance troughs" or holes (see e.g. Chen et al. 1995 and Pfeiffer et
al. 1997). The Hilf et al. (1976) droplet model with schematic shell
corrections did this part reasonably well, although deviations of a factor
of 3 can occur for A=205. The Finite Range Droplet Model with microscopic
shell corrections (M\"oller et al. 1995) and the ETFSI-1 (Aboussir et al. 
1995) are known for their pronounced shell effects, also far from stability.
This leads to the extremely large deviations of up to a factor of 34
and excludes these mass models for a reliable extrapolation into unknown
territory for heavier nuclei. The Hartree-Fock-Bogoliubov calculations with
Skyrme force SkP by Dobaczewsky et al. (1996) are a fully microscopic
self-consistent calculations, where the shell quenching far from stability,
close to the neutron drip-line, results from interactions of loosely bound
neutrons with this neutron continuum. However, the disadvantage is that
these highly advanced and computationally expensive calculations still
assume spherical symmetry for all nuclei. Thus HFB/SkP cannot be a good
overall representation for astrophysical applications yet. We see this
by an up to a factor of 6 overprediction of nuclei in the mass range
203-205. Attempts to improve the global behavior of a mass model like
FRDM with pronounced shell effects by cutting in HFB/SkP at shell closures
have been undertaken by Chen et al. (1995) and are shown here as FRDM/HFB.
We see an improvement in comparison to FRDM alone, but still large deviations.

The best and close agreement is found with ETFSI-Q (Pearson, Nayak, \& Goriely 
1996), the version of ETFSI-1 where a phenomenological quenching of shell 
effects was introduced, combined with a consistent treatment of deformation.
In fact, ETFSI-Q is the only mass model which yields a very good agreement
with $r$-abundances over the whole mass range (not only in the $r$-process
peaks, avoiding large troughs before and after shell closures). This is the
reason why an overall least square fit made sense (not only for the three
peaks, as in case of the other mass models). This is listed as a second
entry ETFSI-Q (lsq.) and changes the coefficient and power in $n_n$ for
the superposition weight $\omega(n_n)$ (see above) to $9.85\times 10^6$
and -0.296.
For the isotopic abundances in the range A=125--209, the deviations 
from solar $r$-abundances yield an uncertainty characterized by a reduced 
$\chi^2$=0.76,
which indicates that the predicted values are not significantly statistically
different from the observed solar system values, with deviations typically
in the 10\% range. In the 203-205 mass region the agreement is slightly
worse, showing on average more a 20\% deviation, about a factor of 2 larger
than the ETFSI-Q entry, which resulted from a fit to the more certain r-process
peak
regions alone. This can also be seen in Figure 1, where both results are
plotted (dashed and solid).

The $r$-process contributions for $^{206-208}$Pb and $^{209}$Bi are dominated 
by alpha-decay chains from heavier nuclei up to A=255. Therefore, they 
represent a major test for the accuracy of our predictions for the heaviest
$r$-process nuclei in the actinide region. The solar $r$-process abundances
for $^{206}$Pb and $^{207}$Pb are quite well understood
due to precision neutron capture cross sections which enter $s$-process
studies (15-19\% accuracy, K\"appeler et al. 1989). This is somewhat worse
for $^{209}$Bi (43\%), where the uncertainty range for the $r$-contribution
lies within 64 and 92\% of the solar abundance (K\"appeler et al. 1989;
Beer, Corvi, \& Mutti 1997). For $^{208}$Pb, which is
dominated by the quite uncertain strong $s$-process component,
K\"appeler et al. (1989) determined the $s$-contribution to be
93.5\%, Beer et al. (1997) 89\%, while stellar site related calculations
(Gallino 1998, private communication) seem to be able to contribute as
much as 97\%. This leaves for the $r$-process contribution between 3
and 11\%, i.e. a mean value of 7\% with a 57\% uncertainty.
Summarizing this part on the $r$-process abundances, suggests that
A=206 and 207 provide clean $r$-process abundances with 15-19\% accuracy,
while A=208 and 209 have to be taken with some caution. This means that
two of the four alpha-decay chain abundances from heavier $r$-process nuclei
are well constrained and two other alpha-decay chains can be taken as a
consistency check. We have not listed here the analysis of Goriely (1997),
because it suffers from uncertainties at the edge of fitting intervals
(discussed above) and does not take into account the knowledge that the
$r$-process contributions behave smoothly as a function of isotopic mass
(see K\"appeler et al. 1989). The latter is observed for the global mass
range of heavy $r$-process nuclei beyond the A=80 peak and not expected
to break down for masses beyond A=200, which also leads to smooth variations
when distributed over the decay products of four alpha-decay chains. 

When comparing in Table 1 the solar $r$-abundances of the nuclei 206 to 209 
with predictions from different mass models, we obtain an additional and
crucial test for their application in predicting actinide abundances.
While the Hilf model was behaving reasonably in the 203-205 mass region,
which tested mostly shell effects, it shows deviations up to a factor of 6
for these Pb and Bi abundances and is therefore not usable for reliable
actinide predictions. The mass models FRDM and ETFSI-1, which already
showed problems with shell effects for 203-205, also lead to an appreciable
underprediction for 206-209 (up to a factor of 3). The hybrid model
FRDM+HFB shows improved results, but deviations by a factor of 2 are
still observed. This improves further with HFB/SkP, but the known deficiency
of this purely spherical approach (plus its problems for 203-205) leave
some doubts in its predictability. Only ETFSI-Q and ETFSI-Q (lsq) show an
overall good performance. ETFSI-Q is actually better with a typical
10-20\% accuracy. This leaves us with the conclusion that applications
to actinide predictions make sense in the sequence
ETFSI-Q and  ETFSI-Q(lsq) (which are consistent with the solar $r$-abundances
within the given errors), with the possible inclusion of FRDM/HFB and maybe
HFB/SkP. The other models listed in Table 1 did not pass the test to
reproduce the "actinide decay products" A=206-209.
The fact that we can reproduce 
the solar Pb and Bi {\it r}-abundances,
produced mainly by four different alpha-decay chains from 
the heavy mass
    region beyond A = 209 up to about 256, where (apart from Th and U) no 
    ``observables'' exist to tune our fits, 
gives us confidence that we are correctly predicting the
radioactive actinide abundance values as well.

In fact, our results demonstrate that all possible observables related to 
actinide abundances are satisfied with an extension of components up to 
$n_n$ of $3\times 10^{27}$, obeying the same power laws for $\omega$ and
$\tau$
as before. (At $3\times 10^{27}$ the abundance pattern is converged,
a further extension to higher neutron densities has no effect, as the
weights become so small that such components do not contribute.) This
has the advantage that no additional parameter (like an upper limit for
$n_n$ components) is required.
The remaining
abundances of $^{232}$Th ($\tau_{1/2}$=$1.405\times 10^{10}$y),
$^{235}$U ($\tau_{1/2}$=$7.038\times 10^8$y) and
$^{238}$U ($\tau_{1/2}$=$4.468\times 10^9$y) will then, after their
production, decay on their long decay-times. The abundances of
$^{232}$Th, $^{235}$U, $^{238}$U, and $^{244}$Pu have already been
shown in Table 1 and 2 of Pfeiffer et al. (1997) for most of the same
mass models and will not be repeated here, with the exception of $^{232}$Th
which is used later for the Th/Eu chronometer.

Beta-delayed fission during the {\it r}-process was neglected in the 
calculations, because earlier investigations (Thielemann, Metzinger, 
\& Klapdor 1983; Cowan et al. 1987; Thielemann, Cameron, \& Cowan 1989;
Cowan, Thielemann, \& Truran 1991),
when constrained by experimental data (Hoff 1986, 1987), have shown
that to first order fission is unimportant.
After the end of the {\it r}-process we considered spontaneous fission for 
nuclei beyond A = 256 (i.e., $^{256}$Cf) and the few known nuclei undergoing
spontaneous fission for A $<$ 256.
Their fission products will make no contribution to the mass range A=232-255 
(which includes the Th and U isotopes). Therefore, only nuclei with
A$\le$255 were included in the decay chains.
We want to state again that the correct prediction for the {\it r-}process 
contribution to the A=206, 207, 208, and 209 nuclei,
produced mostly by alpha-decay chains starting 
in the mass region A=232-255,
gives a strong indication that the total amount of matter in the
A=232-255 mass range is well predicted within this model. 
While extrapolations always involve some risk, the pleasing agreement
noted here for four different alpha-decay chains suggests that our 
calculations also have good predictive power 
in this region of the nuclear chart.

To summarize this section, the excellent fit of our calculations to the
solar system isotopic abundance distribution for A$>$130 suggests that 
one type of astrophysical event has been responsible for {\it r-}process 
production throughout this mass range.
We cannot yet identify what type of event that was, but we expect that
the processed matter experienced a smoothly varying set of thermodynamic
and neutron density conditions, which led to such a good agreement with
solar $r$-abundances.
We caution the reader that our insufficient knowledge of the properties 
of nuclei far from stability could contribute to uncertainties in
the abundance predictions for the chronometer elements Th and U, but
by performing numerous tests, we examined the nuclear physics input which
best reproduces all observables (including also the A=206-209 nuclei
that most constrain the interesting radioactive mass region). These tests 
(to be
discussed further in \S 4.2) also reveal a clearly indicated error range.

\section{Abundance Comparisons} 

We next compare our theoretical {\it r}-process calculations with solar
system {\it r}-process and stellar
elemental abundance distributions. 
In Figure 2 we consider abundance data for the ultra-metal-poor star
CS~22892-052 ([Fe/H]$ \simeq -$3.1).\footnote
{We adopt the usual spectroscopic notations that
[A/B]~$\equiv$~log$_{\rm 10}$(N$_{\rm A}$/N$_{\rm B}$)$_{\rm star}$~--
log$_{\rm 10}$(N$_{\rm A}$/N$_{\rm B}$)$_{\odot}$, and that
log~$\epsilon$(A)~$\equiv$~log$_{\rm 10}$(N$_{\rm A}$/N$_{\rm H}$)~+~12.0,
for elements A and B. Also, metallicity will be arbitrarily defined
here to be equivalent to the stellar [Fe/H] value.}
The observations from Sneden et al. (1996) are indicated by large filled 
squares, while the small squares are the solar system {\it r}-process
abundances connected by a dashed line. 
The solar elemental abundances are of course sums over the isotopic
abundances, and these are based on deconvolving the total solar 
system abundances (\cite{and82,and89}) into {\it s}- 
and {\it r}-process abundance fractions using the neutron 
capture cross sections of K\"appeler et al. (1989)
and Wisshak et al. (1996; see also  
\cite{sne96} and \cite {bur98} for details).
The solid line in Figure 2 indicates
the predicted abundances from the ETFSI-Q 
calculations. Both the 
solar data and the predicted {\it r}-process abundances have been scaled 
vertically downward  
to account for the much lower metallicity of this star with
respect to the sun. 
It is clear from the figure that the stellar data, the scaled solar
abundances and the theoretical {\it r}-process abundances are coincident,
at least for the stable elements Ba (Z~=~56) and above.

Figure 3 shows abundances for the very metal-poor ([Fe/H] = --2.7) halo
giant HD~115444.
It includes the ground-based data (an average of the observational 
values from Griffin et al. 1982 and Gilroy et al. 1988; see also \cite{sne98}) 
indicated by open squares, while HST data from Sneden et al. (1998) are
indicated by solid squares.
In this figure we also include a new abundance value for thorium 
and an upper limit (as an arrow) on uranium. 
These radioactive elements will be discussed in \S 4.  
While Os was detected with somewhat large error
bars in CS 22892-052, there are much more extensive 3rd {\it r}-process peak
data
available for HD~115444 (due to the HST observations).
Sneden et al. (1998) have shown the similarity between the 
solar system {\it r}-process elemental curve and the data (both
ground-based and space-based) all the way through the 
3rd {\it r}-process peak for this star. Our new 
{\it r}-process calculations (the solid line in the figure)
further confirm this result. 

Figure 4 illustrates abundances for HD~122563,
a star of approximately the same metallicity as HD~115444. 
However, HD~122563 is much less abundant in neutron-capture elements 
than is HD~115444, and only upper limits (indicated by arrows) 
for the 3rd {\it r}-process peak elements were obtained by Sneden
et al. (1998). 
The ground-based data are averages of a number of studies (as
discussed in \cite{sne98}). 
Both the solar system scaled {\it r}-process curve and our new {\it r}-process 
calculations are consistent with the available ground-based data and 
the upper limits for  HD~122563  for elements with Z~$\ge$~56. 

The fourth halo star for which there are extensive data is 
HD~126238. This star with a metallicity of [Fe/H] = --1.7 is more
metal-rich than the previously discussed stars. 
Cowan et al. (1996) and  Sneden et al. (1998) have reported  
HST detections of 
the 3rd {\it r}-process peak elements along with Pb in this star, and 
those data are plotted in Figure 5. 
We again see the similarity in the abundance pattern for the 
heaviest neutron-capture elements in this star and the 
scaled solar {\it r}-process abundances, both 
predicted by our calculations  and by actual solar data.
There is, however, some deviation between the Ba abundance
(a predominantly {\it s}-process element) and the {\it r}-process curves,
which may indicate the onset of Galactic {\it s}-processing 
already at this higher metallicity (see also \cite{cow96}, \cite{sne98}),
which agrees well with chemical evolution expectations.
It is possible that some {\it s-}processing might have contributed also
to the Pb abundance, although our data are not adequate to show such an
enhancement.

The overall {\it r}-process to Fe ratio at low
metallicities is highly variable.
This was first noted by
Gilroy et al. (1988), and is certainly evident in the four stars
considered here. 
Specifically, the [Eu/Fe] ratio is +1.56 in
CS~22892--052 (Sneden et al. 1994),+0.70 in HD~115444 as found in 
this paper, --0.44 in HD~122563 (Gratton \& Sneden 1994) 
and +0.17 in HD~126238 (Gratton \& Sneden 1994).
Burris et al. (1998) will discuss this phenomenon more extensively.
These results demonstrate that individual {\it r}-process events are resolved, 
on time scales probably shorter than Galactic mixing time-scales, but 
that the {\it r}-process abundances in each event are apparently well 
reproduced by a solar {\it r-}composition.

\section {Chronometers}

Actinide chronometers have been used to 
determine Galactic ages by (i) predicting 
$^{232}$Th/$^{238}$U and $^{235}$U/$^{238}$U ratios in {\it r}-process
calculations,
(ii) applying them in Galactic evolution models, which include assumptions 
about the histories of  star formation rates and 
{\it r}-process production, and finally (iii) comparing these
ratios with meteoritic data, which provide the Th/U and U/U ratios at the
formation of the solar system. Low-metallicity stars have the advantage
that one can avoid uncertainties introduced by chemical evolution modeling. 
The metallicities of the halo stars for which neutron-capture element
data have become available range from [Fe/H] = --3 to about --2.
Typical (Galactic chemical) evolution calculations suggest {\it roughly} the
"metallicity-age" relation [Fe/H]=--1 at $10^9$y, --2  at $10^8$y,  and --3 at 
$10^7$y (see e.g., \cite{chi97,tsu97}).
Even if this estimate is uncertain by factors of 2-3,
very low metallicity stars most certainly were born when the Galaxy was
only $10^7-10^8$y old, a tiny fraction of the present age of the Galaxy.
Thus the neutron-capture elements observed in very low metallicity stars were
generated in one or at most a few prior nucleosynthesis episodes.
If several events contributed, the time interval between
these events had to be very short in comparison to
thorium decay ages. Thus, no error is made by simply adding these
contributions, without considering decays, and treating them as one
abundance distribution which undergoes decay until the present time. 
Any prior Galactic evolution can therefore be treated as a single 
{\it r}-process event incorporated into the observed star.

\subsection{New Observations}

Sneden et al. (1996) used the technique described above in deriving a 
thorium-based age for CS~22892-052.
Detailed analyses for other low metallicity stars 
with high neutron-capture element abundance levels
would further strengthen the reliability of such age determinations.
An obvious candidate is HD~115444.
It is about 0.3~dex more metal-rich than CS~22892-052, and has
more moderate overabundances of neutron-capture elements 
(e.g., [Eu/Fe]~$\simeq$~+0.7).
However, it is an easier star for detailed spectroscopy, as it
is much the brighter star (V~=~9.0 against 13.1 for CS~22892-052).
Therefore new ground-based UV and blue spectra of HD~115444, having
high resolution (R~$\simeq$~50,000 to 60,000) and high S/N ($\simeq$~75 
to 200 near \wave{4000}), have been gathered with the McDonald Observatory
2.7m telescope and 2D-coud\'e spectrograph (Tull et al. 1995)
and the Keck~I HIRES (Vogt et al. 1992).
An extensive analysis of the McDonald spectrum of 
this star is in preparation (Westin  et al. 1999);
here we discuss preliminary analysis of only 
the chronometer elements.

In Figure~6 we show small spectral regions surrounding two \nion{Th}{ii} lines
and one \nion{U}{ii} line in our data.
The observed spectra are co-added data from the Keck and McDonald raw 
spectra; significant noise reductions in the spectra were achieved 
by the averaging of the spectra.
Superimposed on each observed spectrum are four synthetic spectra in which
only the abundance of thorium or uranium has been allowed to vary.
To compute these synthetic spectra, we employed the current version of
the line analysis code of Sneden (1973).
We adopted the model atmosphere for HD~115444 derived by Pilachowski et al.
(1996); the atomic/molecular line lists for each spectral region are
described briefly below.

The \nion{Th}{ii} line at \wave{4019.13} in HD~115444 is easily seen in the
observed spectrum displayed in the top panel of Figure~6.
The contaminating absorptions to this complex feature are also apparent.
We used the line list described in Sneden et al. (1996), which was based on
earlier studies of Fran\c{c}ois et al. (1993) and Morell et al. (1992), but 
here we added two $^{\rm 13}$CH lines to the list.
Norris, Ryan, \& Beers (1997) have demonstrated that there are possibly severe 
blending complications to the \nion{Th}{ii} feature caused by these 
$^{\rm 13}$CH lines (see their Figure 1). 
The $^{\rm 13}$CH contamination is greatest in those halo 
giants with either strong CH bands and/or low \carbiso\ ratios.
From analysis of $^{\rm 12}$CH and $^{\rm 13}$CH lines in other spectral 
regions, we estimate \carbiso~$\simeq$~7--8 for HD~115444.  
The overall CH band strength was a free parameter that was adjusted to match
the $^{\rm 13}$CH doublet at $\lambda\lambda$4020.0,4020.2~\AA,
and then the $^{\rm 13}$CH absorption at \wave{4019} was determined by
this band strength, the transition probability ratio between the relevant 
$^{\rm 12}$CH and $^{\rm 13}$CH lines, and the \carbiso\ ratio.
The major remaining contamination of the \nion{Th}{ii} feature is due to 
\nion{Co}{i} (chiefly at \wave{4019.3}); and we altered the assumed Co 
abundance to match this absorption.
Finally, we call attention to the partial \nion{Nd}{ii} blend at 
\wave{4018.84}; its strength in our HD~115444 spectrum is well matched by 
the Nd abundance recommended in earlier analyses of this star (Griffin 
et al. 1982, Gilroy et al. 1988).
With these features accounted for, the remaining absorption near \wave{4019}
is attributed to \nion{Th}{ii}, and we deduced log~$\epsilon$(Th)
=~--2.1~$\pm$~0.1, the uncertainty being a combination of goodness-of-fit,
continuum placement, and blending agent uncertainties.

Having a such good resolution and low noise spectrum of this star,
we conducted a search for other strong \nion{Th}{ii} transitions.
This exercise proved mostly fruitless, yielding only a possible
line at \wave{4086.52} that is more than 4 times weaker than the 
\wave{4019} line.
The observed spectrum given in the middle panel of Figure~6 shows
no obvious \nion{Th}{ii} absorption.
Nevertheless, we constructed a line list for features near the \wave{4086} 
line from Kurucz's extensive atomic and molecular line lists (see
Kurucz 1995a,b for descriptions of these data), and altered the
transition probabilities to provide acceptable matches to the
spectrum of the similar metallicity giant HD~122563; see Westin 
et al. (1999) for a more complete description.
Application of this line list to predict the HD~115444 spectrum yielded
the synthetic spectra in the middle panel of Figure~6.
We could deduce only an upper limit to the thorium abundances here:
log~$\epsilon$(Th)~$\leq$~--2.2.

We tried to identify lines of uranium in the HD~115444 spectrum.
Our brief reconnaissance produced only a potentially detectable \nion{U}{ii}
feature at \wave{3859.58}.
Again, no absorption is obvious in the observed spectrum (bottom panel,
Figure 6).
Using the same procedure as we did for the \nion{Th}{ii} \wave{4086} line,
we created a line list for this spectral region and computed synthetic
spectra for HD~115444.
From the observed/synthetic spectrum comparison, we suggest that
log~$\epsilon$(U)~$\leq$~--2.5.

We also obtained McDonald and Keck~I spectra at very high S/N for the
bright, very metal-poor giant HD~122563.
This star, however, is deficient in neutron-capture elements
([Eu/Fe]~$\simeq$~--0.44).
We expected a weak-to-absent \nion{Th}{ii} \wave{4019} line, and
found none in the observed spectrum.
A synthetic spectrum computation yielded only that
log~$\epsilon$(Th)~$\leq$~--3.1.
A Keck spectrum was obtained for CS~22892-052, and although it
has high resolution, the S/N is low.
Therefore we chose to smooth this spectrum (sacrificing resolution
for S/N) and then to merge it with the CTIO 4m echelle spectrum employed
by Sneden et al. (1996).
We again first derived \carbiso~$\simeq$~16 (in excellent agreement
with the estimate of Norris et al. 1997) from other CH features in
the CS~22892--052 spectrum.
Then analysis of the \wave{4019} blend suggested log~$\epsilon$(Th)
=~--1.6~$\pm$~0.1 (Sneden et al. derived --1.55 from just their CTIO
spectrum, but see also Sneden et al. 1999 for additional abundance
 determinations 
for CS~22892--052.)

A summary of our new abundance results for HD~115444, HD~122563, and
CS~22892-052 is given in Table~2.
We also used synthetic spectrum computations of the strong \nion{Eu}{ii}
lines at $\lambda\lambda$4129,4205~\AA\ to derive new values of
the europium abundances in these stars.
Our results here are in good accord with previous determinations (Griffin
et al. 1982, Gilroy et al. 1982, Sneden et al. 1996, 1998).
We have not obtained spectra of the radioactive elements in
HD~126238, and thus this star will play no part in the age discussion
to follow.

\subsection{Ages}

The agreement between the stable stellar abundances observed in low
metallicity stars and the solar $r$-process abundances, demonstrates that
these
stars experienced only an $r$-process contribution for elements beyond
Ba and that this contribution seems indistinguishable from a solar
$r$-process mix. Thorium and uranium are produced solely in the 
{\it r}-process, but they are unstable and decay on long timescales.
Knowing their production abundances in an $r$-process site and comparing
them with the present observed abundances in a low metallicity star 
allows the stellar age to be determined  
(or the time when the pre-stellar
cloud experienced the last pollution by $r$-process matter). 
For the cases with very low metallicities,
these stars are expected to have formed
not longer than $10^8$~y after the formation of the Galaxy. Therefore,
their ages are, within small uncertainties, identical to the age of the
Galaxy.

With the robust fits between stable solar $r$-abundances and our theoretical 
$r$-process predictions as discussed in \S2 and \S3, we can also make use of 
the abundances of Th and U as predictions for the 
zero decay-age abundances of these radioactive elements.
We have therefore compared our theoretical estimates for the initial 
radioactive abundances with observed abundances in the metal-poor
halo stars. Recall from Figure 1, where the abundances  
were compared before and after
decay, that the calculations match very well  (after decay)  
all the  stable {\it r}-process nuclei (even the heaviest Pb and Bi 
isotopes) in the solar system.
That agreement  strengthens our abundance predictions for the 
radioactive nuclei synthesized in the {\it r}-process.

The predicted Th/Eu ratio at zero decay-age, based upon the ETFSI-Q mass
model 
and a superposition of components (see \S 2) according to a least square fit 
to solar $r$-abundances in the mass range 125-209, is 0.48 or 
log $\epsilon$(Th/Eu)$_0$ = --0.32. 
For the ETFSI-Q model the ratio is 0.546 or log $\epsilon$(Th/Eu)$_0$ =
--0.26.
We compared this value with the observed log $\epsilon$(Th/Eu)$_*$ 
abundance ratios for the three very metal-poor halo stars 
HD~115444, CS~22892-052, and HD~122563 given in Table 2.
Ignoring the case for HD~122563, where we only have an upper limit,
we note that the Th/Eu ratios in the other two stars agree
within observational uncertainties of $\sigma$[log $\epsilon$
(Th/Eu)]=$\pm$0.08. 
(See also Sneden et al. 1999.) 
Averaging their ratios (i.e., using log $\epsilon$ (Th/Eu) 
= --0.63), yields a lower limit for the average age of these very 
low metallicity stars of 13.8 Gyr. This results from the mere
fact that the solar Th/Eu ratio listed in Table 3 (at the formation of the 
solar nebular 4.5 billion years ago) represents a lower limit to the
zero decay-age $r$-process abundances, because Eu is stable and Th (although
constantly produced and ejected into the interstellar medium) is partially
decayed.
Sneden et al. (1996) originally estimated an uncertainty in the
$\epsilon$(Th/Eu) of $\pm$0.08 for CS~22892-052, which yielded an
age uncertainty of about $\pm$3.7~Gyr.
The only error sources considered then were observational/analytical
uncertainties in the abundance determination.
With two stars analyzed here, yielding   
essentially the same Th/Eu
ratios,
we estimate the uncertainties to be 
decreased slightly,
and suggest an ``observational'' age uncertainty for the stellar pair 
to be $\pm$3.5~Gyr.

In Table 3 we also list results with zero decay-age Th/Eu ratios from
our $r$-process calculations of \S 2, making use of a variety of nuclear
mass models far from stability. Our ``best'' choice, ETFSI-Q, yields ages
between
14.5 and 17.1 Gyr. 
Although not as easily obtainable as in the case of the observations,
we also have to investigate the possible error due to the theoretical
abundance
predictions on which this age determination is based. 
We first test the agreement that is typical for stable isotopes
and elements resulting from our abundance predictions with the ETFSI-Q nuclear
input. The stable element Eu entering our age determination (later) has
an observed solar system elemental abundance by number of 0.0907$\pm$0.0032 
(on a scale where N(Si) = 10$^6$).
The different methods of determining the superposition weights
$\omega(n_n)$, either via only adjusting to the $r$-process peaks or
by performing a global fit, led to the two sets of abundance predictions
ETFSI-Q and ETFSI-Q (lsq).
The ETFSI-Q (lsq) predicts an abundance for the element of
0.087, which agrees exactly within the deduced errors for $r$-process
residuals. ETFSI-Q predicts 0.11533, which agrees within roughly 20\%.
For the isotopic abundances in the range A=125--209, the deviations 
from solar $r$-abundances yield an uncertainty characterized by a reduced 
$\chi^2$=0.76,
which indicates that the predicted values are not significantly statistically
different from the observed solar system values.
If instead we compare predicted and observed {\it elemental} abundances 
in the range 56$\leq$Z$\leq$83, we find a reduced $\chi^2$=0.86,
again reflecting good agreement and with deviations typically
in the 10\% range. Thus, making use of the two sets ETFSI-Q and ETFSI-Q (lsq),
which differ in the Th/Eu production ratio by 13\%, would seem to be
sufficient
to cover the uncertainty range.

It is also possible to estimate a possible error range by
making use of other mass models. They have, however, to fulfill the 
test discussed in \S2, based on the correct reproduction of
the A=206-209 abundances, which are dominated by and thus are a measure of
the decay-chain products of the actinides. 
This led to the exclusion of
the mass models of Hilf, FRDM, and ETFSI-1. 
FRDM is listed in Table 3, but the Eu abundance prediction is off by a factor
of more than 3, underlining the previous finding, and therefore making the
age prediction meaningless.
Our conclusion was that applications
to actinide predictions  only make sense within the sequence
ETFSI-Q and  ETFSI-Q(lsq) (which are consistent with the solar $r$-abundances
within the given errors), with the possible inclusion of FRDM/HFB and maybe
HFB/SkP. HFB/SkP, which already showed the worst agreement among this
chosen set, and could not model deformed nuclei, gives an age prediction of
10.2 Gyr. This is smaller than the lower limit obtained with the solar Th/Eu
ratio and can therefore be neglected as well. 
The average of ETFSI-Q and ETFSI-Q (lsq) and FRDM+HFB (which passed the test
of Table 1 to reproduce the A=203-209 r-process abundances) is 15.6 Gyr.
The range 13.8 to 17.1 of Table 3, which includes the (observational) lower
limit from the solar Th/Eu ratio, can be somewhat generously assigned with an
error of 2 Gyr. In general the fit quality of our theoretical predictions for
the isotopic abundances in the range A=125--209 gave a reduced $\chi^2$=0.76,
which corresponds roughly to a 1$\sigma$ error of (1.00--0.76)x100 = 24\%.
When applying this 24\% uncertainty
to the (theoretical) Th/Eu ratio, this yields an age uncertainty of 3 Gyrs.
Thus, 3 Gyr should be a safe assessment of the theoretical uncertainties
involved in the Th/Eu chronometer ratio.

The combined error has to reflect this error, based on the theoretical
uncertainties due to nuclear model predictions far from stability,
and the observational error.
Adding in quadrature the observational age uncertainty (as discussed above) 
to this uncorrelated theoretical uncertainty yields a total age
uncertainty of approximately 4.6 Gyr, i.e. stellar ages of these low
metallicity stars of 15.6 $\pm$ 4.6 Gyr.
The estimated age remains virtually unchanged from our
previous estimates (Cowan et al. 1997, Pfeiffer et al. 1997).

The uranium abundance in HD 115444 is a very weak upper limit, 
and in fact the value may be much lower.
Nevertheless, we can make some age estimate based upon this upper limit.
We note, however, that unlike the Th elemental abundance 
which depends only upon one isotope, the uranium abundance
results from the decay of two radioactive isotopes with 
very different half-lives.
In the ETFSI-Q model the initial elemental abundance is 
dominated by the shorter half-life isotope, $^{235}$U, while
presumably the elemental abundance in the metal-poor stars is
mostly determined by $^{238}$U. Ignoring the initial $^{235}$U production
entirely (this isotope decays in much less than a Hubble time), and 
only comparing the initial time-zero $^{238}$U and the observed upper limit 
in HD~115444 gives a lower limit on the age of this star of $>$ 7.1 Gyr. 
While this age estimate has much observational and theoretical uncertainty,
it is entirely consistent with the value found using the Th abundance.

\section {Discussion and Conclusions}

The abundance comparisons shown in the figures indicate several 
striking results.
First, as noted already by Sneden et al. (1998), for the four metal-poor stars 
for which extensive data have become available, the stellar data are in 
solar {\it r}-process proportions (at least for Z~$\ge$~56). 
While this has been suggested now for a number of years,
this trend is now seen over a wider elemental range, 
including data from the 3rd {\it r}-process peak, than was previously
attainable. Future observations including Pb and Bi data, which are
dominated by alpha decay-chains from actinide nuclei, would give further
support for the method applied in
this paper, where we have emphasized the use of theoretical {\it r}-process 
calculations in order to obtain zero decay-age abundances of Th. 
Furthermore, the fact that all four stars, covering a wide range of 
metallicities ([Fe/H] = --3.1 to --1.7), show the same consistent scaled 
solar {\it r}-process pattern, make other explanations (such as a random 
superposition of varying $r$-process abundances from different r-process
sites with varying conditions, see e.g., Goriely \& Arnould 1997) 
much less likely than a continuous set of conditions reproducing 
solar $r$-abundances in each single site.
For elements with Z $\geq$ 56, the relative elemental {\it r}-process 
abundances appear not to have changed over the history of the Galaxy.
This further  suggests that there is one {\it r}-process site in the Galaxy,
at 
least for those elements.

However, the trend for the lighter neutron-capture elements
with A~$<$~100 is not as clear, but may be explainable 
in terms of the emergence of the weak {\it s}-process with 
increasing metallicity.
The abundance values of Zr and Ge, for example, are approximately the same in 
HD~115444 and HD~122563, two stars of the same metallicity but which show 
large differences in the absolute levels of heavy neutron-capture elements.
The weak {\it s}-process is expected to be metallicity
dependent as a secondary process, but to occur earlier than the main
{\it s}-process component, as it is produced in (core He-burning of) massive
stars rather than (He-shell flashes of) low/intermediate mass stars.
We note, however, that it is unclear how effective the weak {\it s}-process
might be in stars of extremely low metallicity, and other explanations 
may be possible. 
One such possibility may be that there are two {\it r}-process signatures --
one for the lighter and one for the heavier {\it r}-process nuclei --
reflecting two separate sites (Wasserburg, Busso, \& Gallino 1996;
Qian, Vogel, \& Wasserburg 1998). 
(See Wheeler, Cowan, \& Hillebrandt 1998, 
and Freiburghaus et al. 1999 for a discussion of possible second sites.)
It is also conceivable that this region of Sr-Zr could be produced 
by a combination of the {\it r-} and the weak {\it s}-process, as suggested
for the star CS~22892--052 by Cowan et al. (1995).
Clearly more observational and theoretical work will be 
required to understand the production of these lighter n-capture
elements early in the history of the Galaxy.

The detection of thorium in very metal-poor stars (pioneered by
Fran\c{c}ois, Spite, \& Spite 1993) has provided the exciting opportunity of 
directly determining stellar ages (see \cite{sne96}, \cite{cow97}
and \cite{pfe97}).
In this paper we have used the stable stellar and solar data to constrain
the predicted (zero decay-age) values of the radioactive {\it r}-process
nuclei.
With detailed \nion{Th}{ii} analyses now in two very metal-poor stars, 
we have extended our previous age determinations to give an average
age estimate of about 15.6 Gyr for very old halo stars, supporting  
previous age estimates for CS~22892--052 
from Cowan et al. (1997) and Pfeiffer et al. (1997). 
Recent age determinations with the aid of the Re/Os chronometer
(Takahashi et al. 1998) yield very similar ages and uncertainty
ranges. The advantage of much better (by now experimentally) known nuclear
properties and missing observational uncertainties is compensated in the
Re/Os chronometer by
the uncertainties involved in galactic evolution and the temperature history
of matter in stars and the interstellar medium. The latter is missing
here because, after an initial $r$-process injection, only Th decay
has to be accounted for.

It is also encouraging to note that our new age estimate for these two stars
is consistent with recent globular cluster age determinations based upon 
Hipparcos data (see Pont et al. 1998).
We note further the consistency of this age estimate with the recent 
cosmological age estimates, based upon high-redshift supernovae, 
of 14.9 $\pm$ 1.5 Gyr (Perlmutter et al. 1999) and 14.2 $\pm$ 1.7 Gyr  
(Riess et al. 1998).
We caution, however, that there are still several uncertainties
which limit the accuracy of these radioactive age determinations. 
In particular, small changes in the 
abundance determinations can result in large age uncertainties
due to the exponential decay-time dependence. 
The resulting age similarities for the two stars for which data are 
available, however, do lend support to using this technique, which
we emphasize is independent of chemical evolution models.
To refine further and strengthen this technique will require many 
more observations
of the long-lived radioactive elements in other metal-poor halo stars.
Additional theoretical calculations, employing full, dynamic {\it r}-process
calculations,
suitable for astrophysical environments, 
will also be needed to make more accurate stellar and Galactic age
determinations.

\acknowledgments

We thank John Norris and Jim Lawler for helpful conversations, 
and appreciate the expert help of the STScI staff in all aspects 
of the planning and execution of the observations over three HST Cycles.
We also thank an anonymous referee who 
helped us to improve the paper, especially on the treatment of 
the nuclear uncertainties. 
Further thanks go to P. M\"oller for the QRPA code, M. Pearson for the ETFSI-Q
masses, and B. Neas and L. D. Cowan,  
Dept. of Biostatistics and Epidemiology 
at OUHSC, for helpful advice on statistics and
uncertainties.
Support for this work was provided through grants GO-05421, GO-05856, 
and GO-06748 from the Space Science Telescope Institute, which is 
operated by the Association of Universities for Research in Astronomy,
Inc., under NASA contract NAS5-26555. 
Support was also provided by the National Science Foundation 
(AST-9314936 and AST-9618332 to J.J.C.),    
(AST-9315068 and AST-9618364 to C.S.) and (AST-9529454 to
T.C.B.), by the Swiss Nationalfonds 
(grants 20-47252.96  and 2000-53798.98), and by the German 
BMBF (grant 06Mz864) and DFG (grant Kr8065).
J.J.C. and F.-K.T. thank the Institute for Theoretical Physics
at the  University of California at Santa Barbara, 
for partial support from  the National
Science Foundation under Grant No. PHY94-07194. 
This work was completed while J.J.C. was in residence at the University
of Texas at Austin, funded through the University of Oklahoma as a Big XII
Faculty Fellow; we are very grateful to both institutions for
their support of this research.

\begin{deluxetable}{rlllllll}

\tablewidth 6in

\tablenum{1}
\tablecaption{Isotopic r-Abundances in Pb-Region (in $10^{-3}$)}
\tablecolumns{8}
\tablehead{
\colhead{Model}                      &
\colhead{$^{203}$Tl}                       &
\colhead{$^{204}$Hg}                       &
\colhead{$^{205}$Tl}                       &
\colhead{$^{206}$Pb}                       &
\colhead{$^{207}$Pb}                       &
\colhead{$^{208}$Pb}                       &
\colhead{$^{209}$Bi}                     
}
\startdata
          \cutinhead{Solar r-Abundances}
K\"appeler & 12$\pm$6 & 20$\pm$2 & 41$\pm$15 & 223$\pm$43 &
280$\pm$47  & 118$\pm$167 & 93$\pm$12 \nl
Beer      &  19$\pm$6 & 25$\pm$6  & 34$\pm$15 & 
256  & 228  & 198 & 134  \nl
combined  &  16$\pm$8 & 21$\pm$6  & 38$\pm$21 & 
240$\pm$61 & 254$\pm$66  & 158$\pm$236 & 114$\pm$17  \nl
          \cutinhead{Calculations (Fit to 3 r-Process Peaks)}
Hilf    &  \ 7.5 & \ 25.1  & 13.7 & \ 53.5  &
\ 44.7  & \ 36.3  & \ 39.5  \nl
FRDM      &  \ 2.4 &  \ \ 1.3  &  \ 1.2 & \ 77.4  &
\ 81.2   & \ 46.4  & \ 69.6  \nl
ETFSI-1  & \ 3.2 &  \ \ 2.7  &  \ 2.9 & \ 80.3  &
\ 82.2   & \ 69.5  & \ 56.1  \nl
HFB/SkP   & 44.5 & 131.3  & 54.1 & 190.9 &
140.6  & 162.6  & \ 99.0  \nl
FRDM/HFB   & 13.2 &  \ \ 3.6  & \ 2.3 & 112.1 &
114.9  &  \ 73.8  & 101.2  \nl
ETFSI-Q   & 10.7 & \ 20.6  & 19.4 & 199.3  &
184.1  & 170.3  & 129.8  \nl
          \cutinhead{Calculations (Least Square Fit A=125-209)}
ETFSI-Q   &  \ 8.5 & \ 16.4  & 15.4 & 157.9 &
146.2  & 135.2  & 102.9  \nl
\enddata
\end{deluxetable}

\begin{deluxetable}{cccc}

\tablewidth 4.5in

\tablenum{2}
\tablecaption{Neutron-Capture Element Abundances}
\tablecolumns{4}
\tablehead{
\colhead{Transition}                       &
\colhead{log $\epsilon$}                   &
\colhead{log $\epsilon$}                   &
\colhead{log $\epsilon$}                   \nl
\colhead{ }                                &
\colhead{HD 115444}                        &
\colhead{CS 22892-052}                     &
\colhead{HD 122563}
}
\startdata
          \cutinhead{Derived Abundances}
\nion{Th}{ii} \wave{4019} &    --2.1  &      --1.6  &  $<$--3.1 \nl
\nion{Th}{ii} \wave{4086} & $<$--2.2   &    \nodata  &   \nodata \nl
\nion{U}{ii}  \wave{3860} & $<$--2.5   &    \nodata  &   \nodata \nl
\nion{Eu}{ii} $\lambda\lambda$4129,4205~\AA &
                               -1.5   &     --0.9  &    --2.6 \nl
          \cutinhead{Abundance Ratios}
Th/Eu     &     --0.6   &      --0.66   &   $<$--0.5 \nl
U/Eu      &  $<$--1.0   &    \nodata   &    \nodata \nl
Th/U      &  $>$+0.4    &    \nodata   &    \nodata
\enddata

\end{deluxetable}

\begin{deluxetable}{llllr}
\tablewidth 6.in

\tablenum{3}
\tablecaption{Th/Eu Chronometers}
\tablecolumns{5}
\tablehead{
\colhead{Model}                       &
\colhead{$_{90}$Th}                   &
\colhead{$_{63}$Eu}                   &
\colhead{$\frac{Th}{Eu}$$\vert_{model}$}  &
\colhead{Age [Gyrs]}                  
}
\startdata
    Solar       &  0.042     & 0.09    &  0.463  & 13.8 \\
    \hline
    FRDM        &  0.04280   & 0.02420 &  1.7695 & 41.0 \\ 
    ETFSI-1     &  0.02949   & 0.06041 &  0.4881 & 14.9 \\
    HFB/SkP     &  0.01991   & 0.05134 &  0.3879 & 10.2 \\
    FRDM+HFB    &  0.03449   & 0.06958 &  0.4957 & 15.2 \\
    ETFSI-Q     &  0.06292   & 0.11533 &  0.5456 & 17.1 \\
    ETFSI-Q(lsq)&  0.04222   & 0.08788 &  0.4804 & 14.5 \\
\enddata
\end{deluxetable}


\clearpage
\pagestyle{empty}

\noindent{\bf Figure Captions:}

\figcaption{Comparison of theoretical abundances 
prior to and after beta- and alpha-decay 
with solar {\it r}-process abundances (small filled circles).
For A$>$206 two superposition weights $\omega(n_n)$ have been
utilized, obtained from fitting the $r$-process abundance peaks 
(ETFSI-Q, dashed) and from a global abundance fit in the mass region 
A=125-209 (ETFSI-Q lsq., solid).
The solid vertical lines show the calculated abundance range
within ETFSI-Q and ETFSI-Q (lsq) predictions for the
nuclei $^{232}$Th, $^{235}$U, $^{238}$U and $^{244}$Pu.
(See text for discussion.)
\label{fig1}}

\figcaption{Comparison of observed abundances (large filled squares) 
from CS~22892-052 and solar{\it r}-process abundances (small filled circles) 
joined by a dashed line 
with a theoretical {\it r}-process abundance distribution denoted by 
a solid line. (See text for
discussion.)
\label{fig2}}

\figcaption{
An abundance comparison between the neutron-capture elements in
HD~115444 and a theoretical {\it r}-process  (solid line) and a solar system
{\it r}-process (dashed line) abundance distribution.
Ground-based data (from various sources, see text for discussion) are
indicated by open squares, while HST data from Sneden et al. (1998) are
indicated by
solid squares.
\label{fig3}}

\figcaption{
An abundance comparison for HD~122563 in the same style as that of Figure 3.
Upper limits for Os and Pt are indicated by crosses with attached arrows.
\label{fig4}}

\figcaption{
An abundance comparison for HD~126238 in the same style as that of Figure 3.
\label{fig5}}

\figcaption{
Observed and synthetic spectra of thorium and uranium transitions in
the spectrum of HD~115444.
See text for discussion of the observational data and the synthetic
spectrum computations.
\label{fig6}}

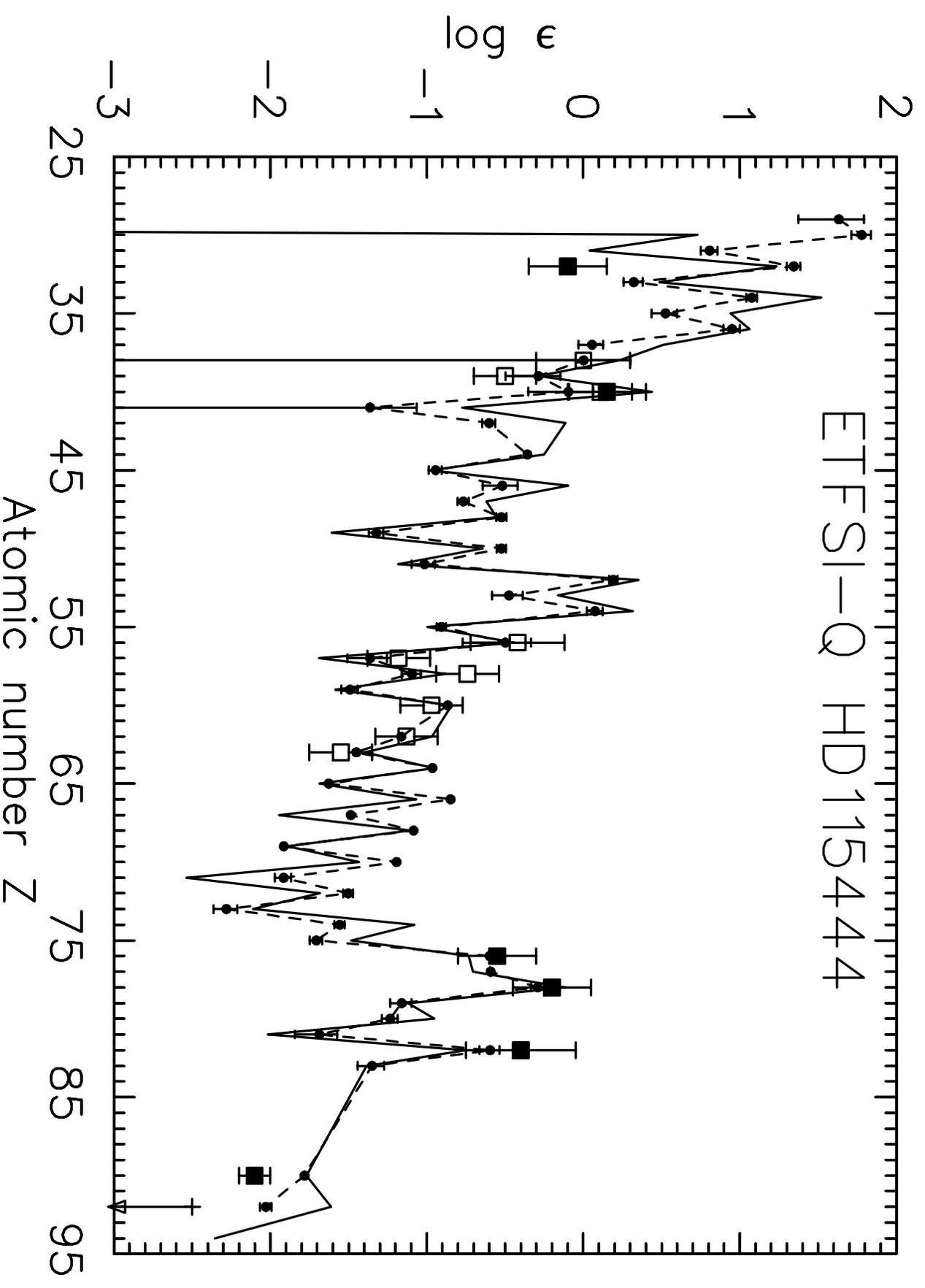
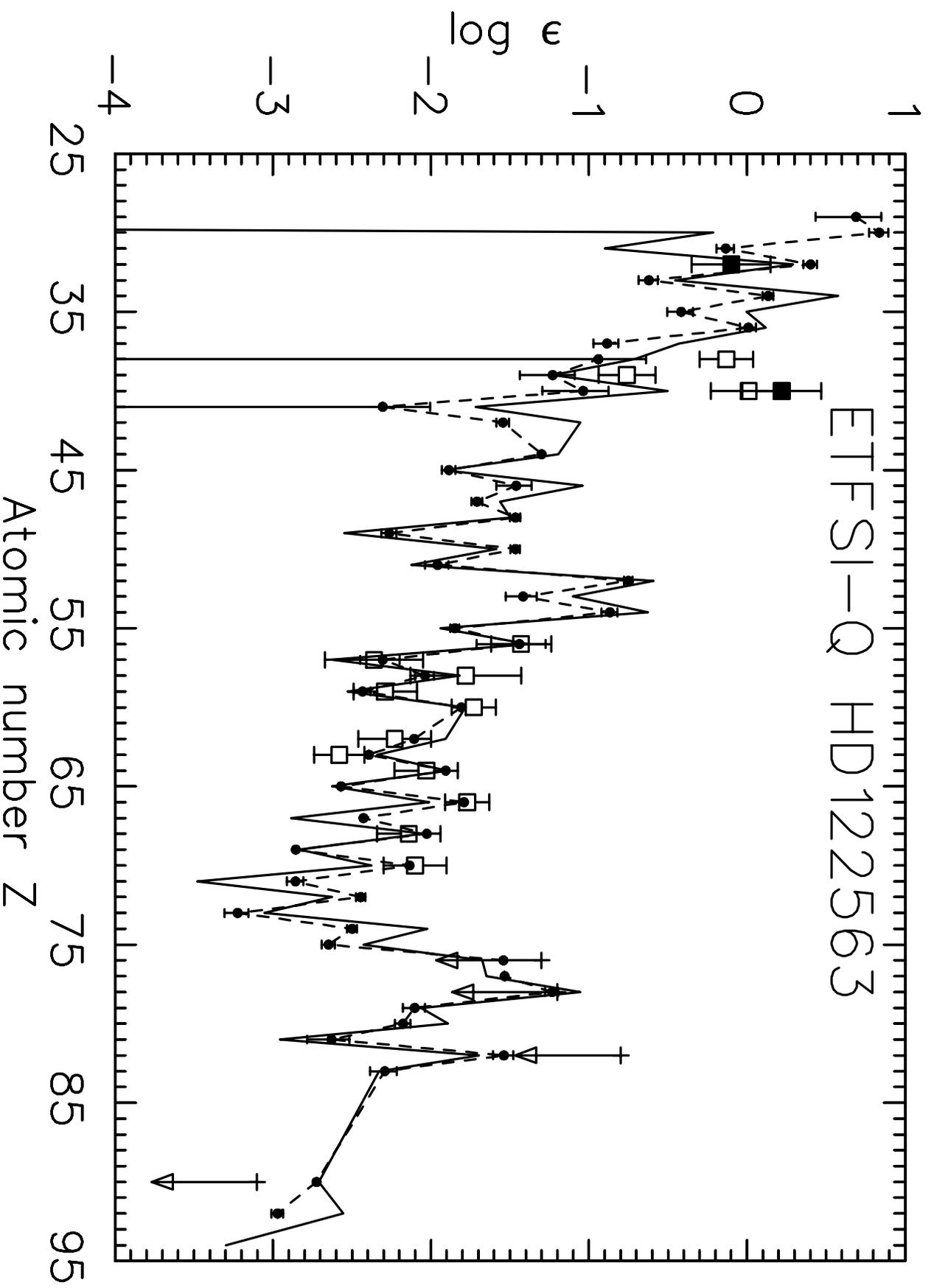
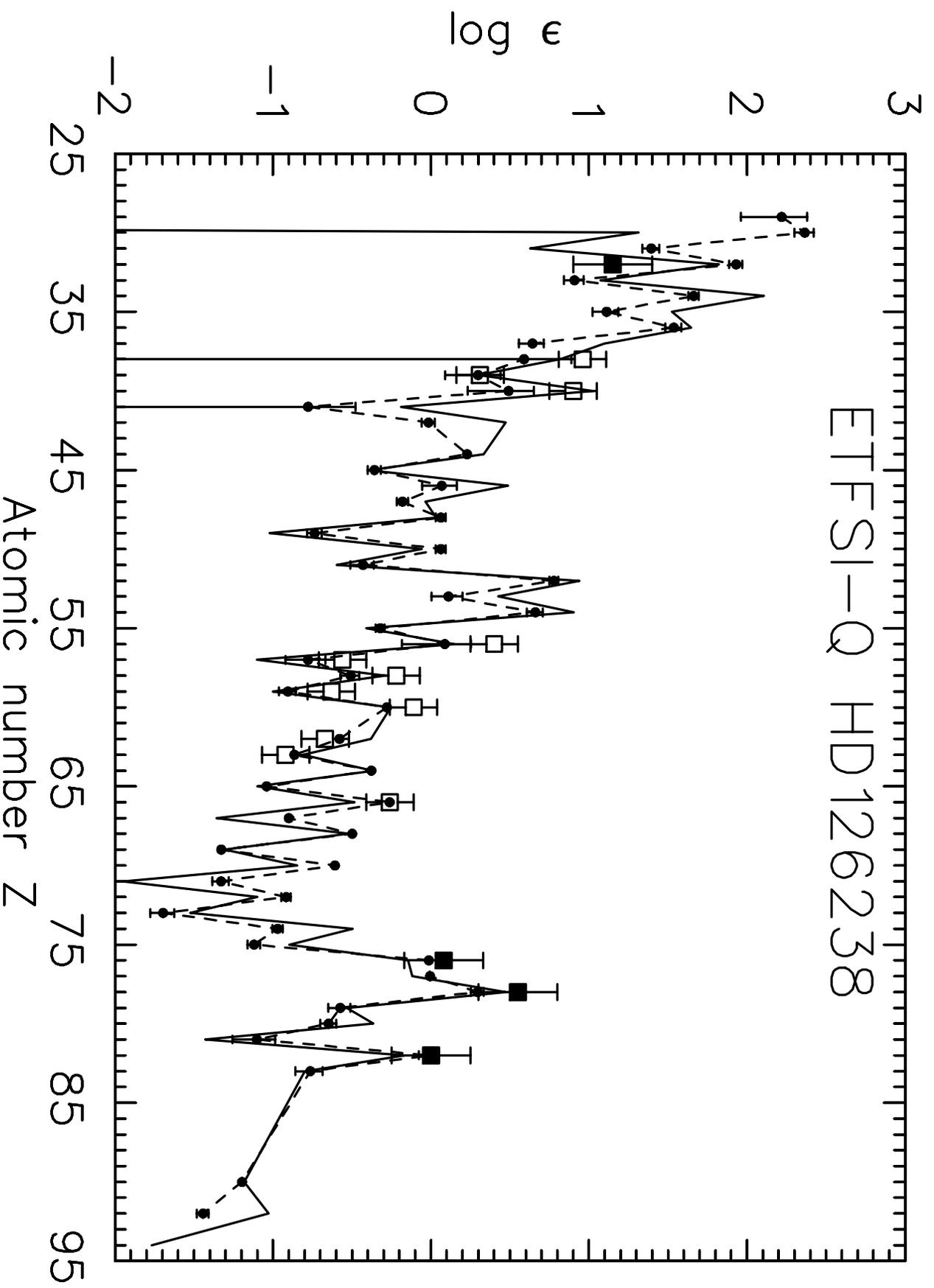
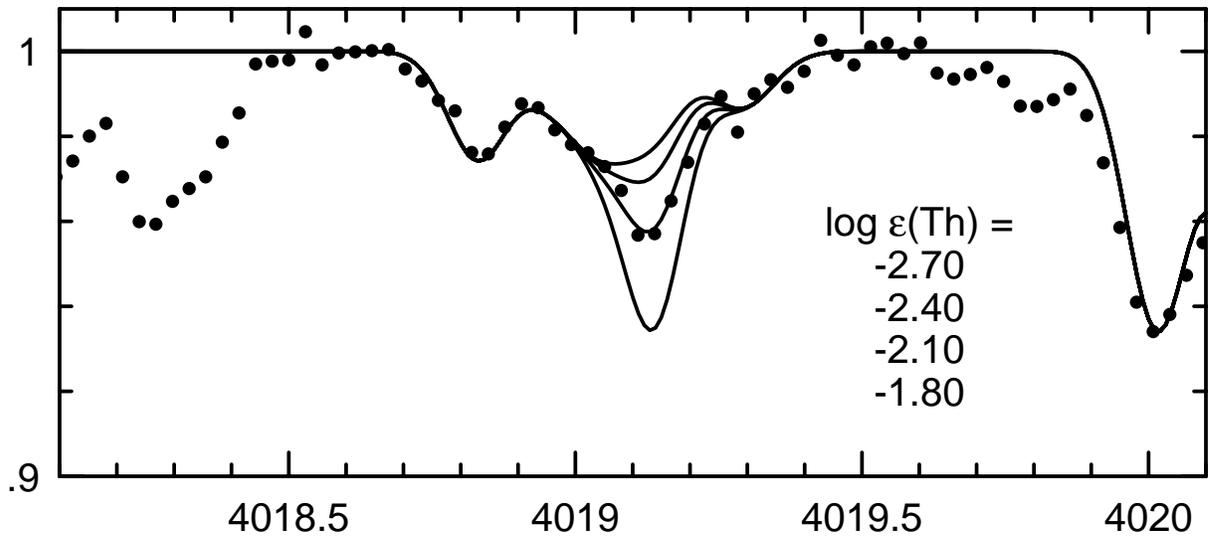
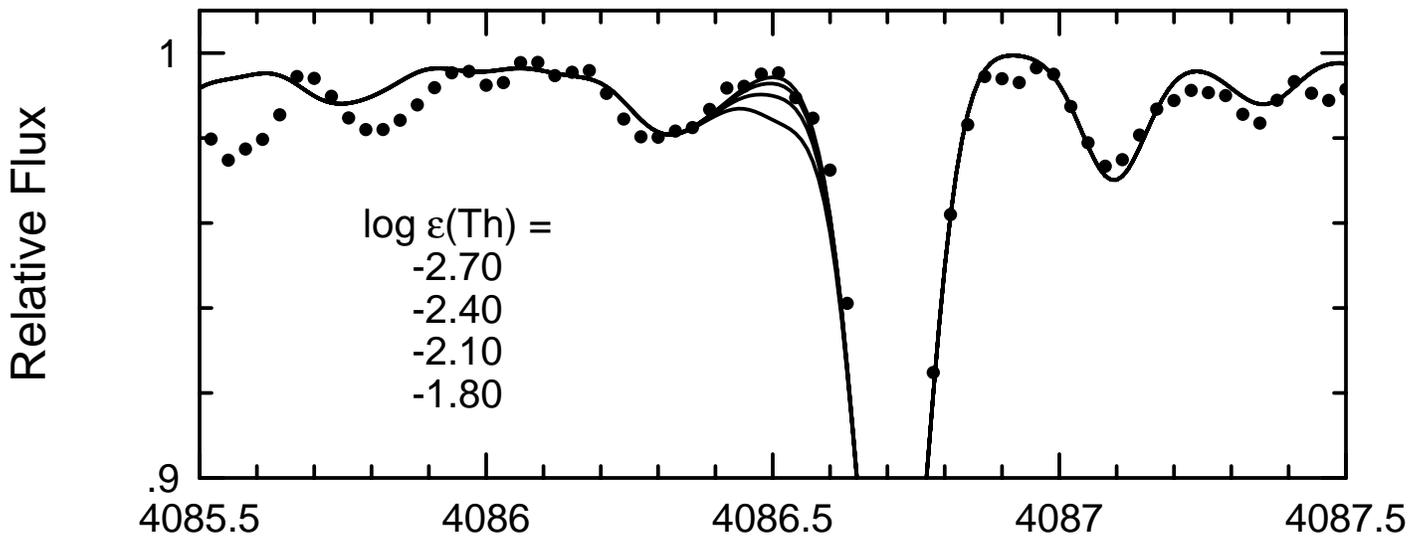
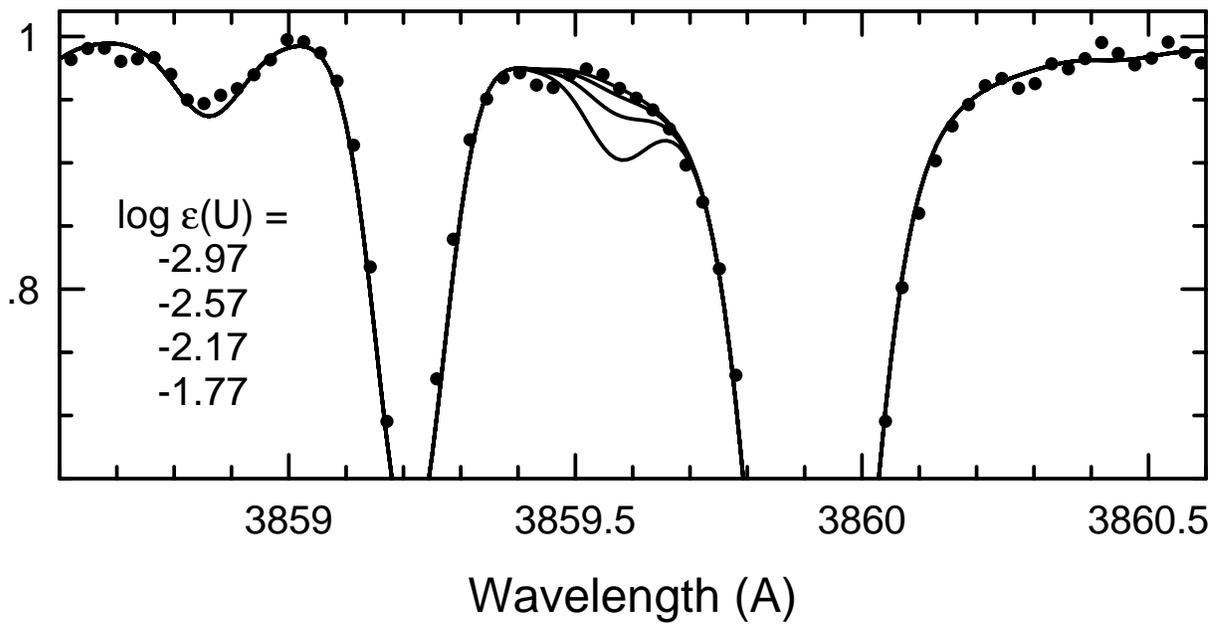
\end{document}